\documentclass[a4paper,10pt]{article}

\usepackage[centertags]{amsmath}
\usepackage{amsmath}

\usepackage{amsfonts}
\usepackage{amssymb}
\usepackage{amsthm}
\usepackage{newlfont}
\usepackage{bbm}

\theoremstyle{plain}
  \newtheorem{theorem}{Theorem}[section]

  \newtheorem{lemma}[theorem]{Lemma}
\theoremstyle{definition}
  \newtheorem{definition}[section]{Definition}
  
\theoremstyle{remark}
  \newtheorem{remark}[theorem]{Remark}

\numberwithin{equation}{section}

  \let\de=\delta 
  \let\ga=\gamma 
 \let\la=\lambda \let\om=\omega

 \let\Ga=\Gamma \let\La=\Lambda

\newcommand{\caB}{{\mathcal B}}
\newcommand{\caC}{{\mathcal C}}
\newcommand{\caD}{{\mathcal D}}
\newcommand{\caE}{{\mathcal E}}
\newcommand{\caF}{{\mathcal F}}

\newcommand{\caH}{{\mathcal H}}

\newcommand{\caN}{{\mathcal N}}

\newcommand{\caZ}{{\mathcal Z}}

\newcommand{\bbC}{{\mathbb C}}

\newcommand{\bbN}{{\mathbb N}}

\newcommand{\bbR}{{\mathbb R}}

\newcommand{\opunit}{\text{1}\kern-0.22em\text{l}}

\newcommand{\frh}{{\mathfrak h}}

\newcommand{\ra}{\rightarrow}

\newcommand{\Span}{{\mathrm{Span}}}
\newcommand{\dyn}{{\mathrm{dyn}}}
\newcommand{\st}{{\mathrm{st}}}
\newcommand{\e}{{\mathrm e}}
\renewcommand{\i}{{\mathrm i}}
\renewcommand{\d}{{\mathrm d}}

\newcommand{\res}{{\mathrm R}}

\newcommand{\beq}{ \begin{equation} }
\newcommand{\eeq}{ \end{equation} }
\newcommand{\bet}{ \begin{theorem} }
\newcommand{\eet}{ \end{theorem} }
\renewcommand{\sp}{{\mathrm{sp}}}
\renewcommand{\dim}{\mathrm{dim}}

\newcommand{\Dom}{\mathrm{Dom}}
\newcommand{\Ran}{\mathrm{Ran}}
\newcommand{\off}{\mathrm{o}}
\newcommand{\ren}{{\mathrm{ren}}}

            \newcounter{smallarabics}
\newenvironment{arabicenumerate}
{\begin{list}{{\normalfont\textrm{(\arabic{smallarabics})}}}
  {\usecounter{smallarabics}\setlength{\itemindent}{0cm}
  \setlength{\leftmargin}{5ex}\setlength{\labelwidth}{4ex}
  \setlength{\topsep}{0.75\parsep}\setlength{\partopsep}{0ex}
   \setlength{\itemsep}{0ex}}}
{\end{list}}

\newcounter{smallroman}
\newenvironment{romanenumerate}
{\begin{list}{{\normalfont\textrm{(\roman{smallroman})}}}
  {\usecounter{smallroman}\setlength{\itemindent}{0cm}
  \setlength{\leftmargin}{5ex}\setlength{\labelwidth}{4ex}
  \setlength{\topsep}{0.75\parsep}\setlength{\partopsep}{0ex}
   \setlength{\itemsep}{0ex}}}
{\end{list}}

\newcommand{\ben}{\begin{arabicenumerate}}
\newcommand{\een}{\end{arabicenumerate}}

\renewcommand\bigoplus{\mathop{\oplus}\limits}
\renewcommand{\Re}{\mathrm{Re}}
\renewcommand{\Im}{\mathrm{Im}}

\begin{document}

\begin{center}
\noindent{\large \bf Extended Weak Coupling Limit\\
for  Friedrichs Hamiltonians} \\

\vspace{15pt}

{\bf Jan Derezi\'{n}ski}
\\
 Department of Mathematical Methods in Physics \\
Warsaw University \\
 Ho\.{z}a 74, 00-682, Warszawa, Poland\\
email: {\tt jan.derezinski@fuw.edu.pl}\\
\vspace{10pt}
{\bf Wojciech De
Roeck} \\
Universiteit Antwerpen \\
Instituut voor Theoretische
Fysica, K.U.Leuven,\\
 Belgium \\
 email:
{\tt wojciech.deroeck@fys.kuleuven.be}\\

\end{center}

\vspace{20pt} \footnotesize{ \noindent {\bf Abstract: } We study a
class of self-adjoint operators defined on the direct sum of two
Hilbert spaces: a finite dimensional one called sometimes a
``small subsystem'' and an infinite dimensional one  -- a
``reservoir''. The operator, which we call a ``Friedrichs
Hamiltonian'', has a small coupling constant in front of its
off-diagonal term. It is well known that, under some conditions, in
the weak coupling  limit the appropriately  rescaled evolution in
the interaction picture converges to a contractive semigroup when
restricted to the subsystem. We show that in this model, the
properly renormalized and rescaled evolution converges on the
whole space to a new unitary evolution, which is a dilation of the
above mentioned semigroup. Similar results have been studied
before \cite{AFL} in more complicated models under the name of
 ``Stochastic Limit".}

\vspace{5pt}

 \vspace{20pt} \normalsize
\section{Introduction}
\subsection{Weak coupling limit}

The weak coupling limit is often
 invoked to justify
various approximations in quantum physics, at least since \cite{VH}.
It  involves a dynamics depending on a small
coupling constant $\lambda$. One assumes that
 \beq \la \searrow 0,\ \, t \ra \infty, \quad \la^2t \textrm{
fixed .} \eeq
Usually one
separates the system into two parts: a ``small subsystem'' and a ``reservoir''.
The long
cumulative effect of the reservoir on the small system
 can in this  limit lead  to a   Markovian dynamics (i.e., a dynamics given
 by a semigroup).

There exists a large literature devoted to the weak coupling limit reduced to
 the small subsystem.
 It
 was first
 put  on a rigourous
footing by Davies \cite{Da1}.
The setup considered by Davies, in its abstract version, consists of a dynamics
 generated by $H_\lambda:=H_0+\lambda W$, a projection $P$ commuting with
 $H_0$ and such that $PH_0P=0$. Davies proved that under appropriate
 assumptions there exists the limit of the dynamics in the interaction picture
 restricted from the left and right by
$P$, and this limit
 is a semigroup on $\Ran P$.
 Perhaps, it would be appropriate to call
the weak coupling limit reduced to the small
 subsystem the
``Davies limit''.  Another name which one can use is the ``reduced weak
 coupling limit''. (In the literature the names ``weak coupling'' and ``van
Hove limit'' are used -- both are  rather imprecise, the latter name is
especially ambiguous, since it is also used for a completely different concept
 in statistical physics).

Davies and a number of other authors gave
applications of the above idea to physically interesting situations describing
a dynamics of a composite quantum system, where $P$ is a conditional
expectation onto the small system and the resulting semigroup is completely
positive. Note, however, that the reduced weak coupling limit is an interesting
mathematical phenomenon also in its more general version.

Some authors point out that it should be possible to use the idea of the weak
coupling limit not just for the dynamics restricted to the small system,
 but for the whole system as well.
In \cite{AFL}, Accardi, Frigerio and Lu
argue that in an appropriate limit, the full unreduced dynamics of a quantum
system converges to a 1-parameter group of $*$-automorphisms, which is a
dilation of the completely positive semigroup obtained by Davies. They call it
the ``stochastic limit''. We prefer to call it the ``extended weak coupling
limit'', since in itself this concept does not have to involve ``stochasticity''.

We believe that the above idea is interesting and worth exploring. In our next
paper \cite{DD} we would like to present our version of the
 extended weak coupling 
 limit applied to 
quantum systems, with some improvements as compared to
\cite{AFL}. 
In particular, we believe that the approach of \cite{DD}  proposes
a more satisfactory kind of convergence (strong*) than that of \cite{AFL}
(convergence of correlation functions) and the proofs of \cite{DD} are  
considerably simpler than those of \cite{AFL}.

\subsection{Weak coupling limit for Friedrichs Hamiltoninas}

In the present paper
we present  results of the same
flavour for a class of simple
operators on a Hilbert space, which we call Friedrichs Hamiltonians.
We will show that for  Friedrichs Hamiltonians the idea of the
extended weak coupling
 limit works very well and yields in a rather natural fashion a
unitary dilation  of the semigroup $\Lambda_t$.

By a ``Friedrichs Hamiltonian'' we mean a self-adjoint operator $H_\lambda$
on a Hilbert space $\caH=\caE\oplus\caH_R$ given by the expression
\begin{equation}
H_\lambda:=\left[\begin{array}{cc}E&\lambda V\\\lambda
    V^*&H_{\res}\end{array}\right] ,\label{gene}\end{equation}
where $E$ is a self-adjoint
 operator on the space $\caE$, $V\in \caB(\caE,\caH_{\res})$ and
$H_{\res}$ is a self-adjoint operator on  $\caH_{\res}$. We will assume that $\caE$ is
finite-dimensional. The subscript ${\res}$ stands for the ``reservoir''.


The Friedrichs model, often under other names such as the Wigner-Weisskopf
atom, is frequently used as a toy model in mathematical physics.
In particular, one often considers its second
quantization on the bosonic or fermionic
 Fock space.
 (Note that the latter  is extensively discussed in
\cite{AJPP}).

For a large class of Friedrichs
Hamiltonians, it is easy to prove that the reduced weak coupling
 limit exists. In this case,
the reduced weak coupling
 limit says that under appropriate assumptions the following limit exists:
\begin{equation}
\lim_{\lambda\searrow0}\e^{\i t E/\lambda^2}1_\caE
\e^{-\i tH_\lambda/\lambda^2}1_\caE=:\Lambda_t,\label{limi}\end{equation}
 and $\Lambda_t$ is a contractive semigroup on $\caE$.

By enlarging the space $\caE$ to a larger Hilbert space
$\caZ=\caE\oplus\caZ_{\res}$, one can construct a
dilation of $\Lambda_t$. This means, a unitary group $\e^{-\i tZ}$ such that
\[1_\caE\e^{-\i tZ}1_\caE=\Lambda_t.\]
The operator $Z$ is actually another example of a Friedrichs Hamiltonian.
 We devote
 Section \ref{sec: dilation} to the construction of a dilation of a
 contractive semigroup that is well adapted to the analysis of the weak
 coupling limit.
 Note that  this construction is quite different from  the
usual one  due to Foias and Nagy \cite{NF}.
Even though it can be found in many
disguises in the literature, we have never seen
a systematic description of some of its curious properties. Therefore, in
Section
\ref{sec: dilation} we devote some space to study this construction.
Note, in particular, that
$Z$ is an example of a
 Friedrichs Hamiltonian  whose definition requires a
``renormalization'' in the terminology of \cite{DF2}.

The main results of our paper are described in Section \ref{stoch}.
We  start  from a rather arbitrary Friedrichs Hamiltonian.
First we describe its Davies limit.
Then  we show that for an appropriate ``scaling operator'' $J_\lambda$
and a ``renormalizing operator'' $Z_\ren$
\[\lim_{\lambda\searrow0}\e^{\i t\lambda^{-2}Z_\ren}J_\lambda^* \e^{-\i
  t\lambda^{-2}H_\lambda}J_\lambda=\e^{-\i tZ}.\]
  It is  this convergence of the dynamics
to a dilation of the semigroup $\Lambda_t$ that
 we call "extended weak coupling limit".
 Following \cite{DF1,DF3}, we will give two versions of
these results: stationary and time-dependent.

Note that the Davies limit follows from the extended weak coupling
 limit, since
\begin{equation}
  1_{\caE} \e^{\i
\la^{-2}tZ_{\mathrm{ren}}}J^{*}_{\la} \e^{-\i
\la^{-2}tH_{\la}}J_{\la} 1_{\caE} = 1_{\caE} \e^{\i \la^{-2}t
{E}} \e^{-\i \la^{-2}t H_{\la}} 1_{\caE}.
\end{equation}

\subsection{The case of 1-dimensional $\caE$}

The main idea of the extended weak coupling
 limit can be explained already in the case of a
one-dimensional small Hilbert space $\caE$.
If $E$ has more than one eigenvalue, which is possible if $\dim \caE\geq2$,
then the extended weak coupling
 limit is more complicated to formulate and prove, which
tends to obscure the whole picture.
Therefore, in this subsection, we describe the main idea of our result in the
case $\dim\caE=1$.

Let $\caE=\bbC$ and $\caH_{\res} = L^2(\bbR)$. Let $e \in \bbR$
and let $\omega$ be a  function on $\bbR$.
 Assume that there is a unique
$\hat{e} := \omega^{-1}(e)$. Let
$\omega$ also stand for the corresponding  multiplication operator on
$\caH_{\res} $ respectively.
 Fix a function $v \in
L^2(\bbR)$ and denote by $\langle v|$ the operator in
$\caB(\caH_{\res},\caE)$ which acts as $\langle v|  := \langle v|f
 \rangle\in \caE$ and let $|v \rangle :=(\langle v|)^*$.
Consider the following Hamiltonian on  $\caE \oplus \caH_{\res} $:
\begin{equation}H_\lambda:=\left[\begin{array}{cc}e&\lambda \langle v|\\\lambda
   |v\rangle &\omega\end{array}\right] ,\label{frie}\end{equation}
(Note that in the literature the name ``Friedrichs Hamiltonian'' is usually
reserved for an operator of the form (\ref{frie}). Operators of the form
(\ref{gene}), should be perhaps called ``generalized Friedrichs
Hamiltonians'').


 The weak coupling limit in this model
simply states that, under some mild assumptions,
 \beq \label{toy result}
 \lim_{\la \downarrow 0} 1_{\caE}
\e^{-\i \la^{-2}t (H_{\la}-e)} 1_{\caE}=\e^{-\i \gamma t},
\end{equation}
where $1_{\caE}$ is the orthogonal projection on 
$\caE$,
\beq
\ga:= \mathrm{P}\int_{\bbR} \d x \, \frac{ v^*(x)v(x)}{\om(x)-e} +\i \pi
    v^*(\hat{e})v(\hat{e}),
\eeq
and $\mathrm{P}\frac1x$ is the principal value of $\frac1x$. 


$\e^{-\i \gamma t}$ is a contractive semigroup on $\caE$. It can be dilated to
a unitary group $\e^{-\i tZ}$ on the Hilbert space
on $\caE \oplus \caH_{\res}$. The generator of the dilating group can be
formally written in the form of a Friedrichs Hamiltonian as
\begin{equation} \label{intuitive G}
Z:=\left[\begin{array}{cc}\Re\gamma
&v(\hat e) \langle \mathbf{1}|\\ v(\hat e)
   |\mathbf{1}\rangle &\omega'(\hat e) x\end{array}\right] .\end{equation}
 $\omega'(\hat{e})x \in \bbR$ is the new multiplication operator on
$\caH_{\res}=L^2(\bbR)$.
$\mathbf{1}$ is the constant function with value $1$ which is of
course not an element of $L^2(\bbR)$. Because of this
(\ref{intuitive G})  does not make  sense as an operator.
Nevertheless, one can give it a precise meaning, e.g. by constructing its
resolvent or its unitary group, or by  imposing a cutoff and taking it away (see e.g. \cite{DF2} and \cite{Ku}).


To state the extended weak coupling limit, we need the unitary rescaling
operator $J_{\la} \in \caB(L^2(\bbR))$ defined as
\begin{equation} (J_{\la}
f)(x)=\frac{1}{\la}f(\frac{x-\hat{e}}{\la^2}), \qquad f \in
L^2(\bbR). \end{equation}  If the function $v$ is sufficiently
regular in $\hat{e}$, we show the following results:
 \ben
 \item{Theorem \ref{thm: stationary}:   the rescaled resolvent $J^*_{\la}(z-\la^{-2}(H_{\la}-e))^{-1}J_{\la}$
 converges in norm to $(z-{Z} )^{-1}$
};
\item{Theorem \ref{thm: dynamic}: the rescaled unitary family $J^*_{\la} \e^{-
    \i t\la^{-2} t
(H_{\la}-e)}J_{\la}$
 converges strongly to $\e^{- \i t {Z}}$.  }
\een


\subsection{Notation}
 We will often make the following abuse of notation.
If $\caH_{0}$ is a closed subspace of a Hilbert space $\caH$, $
 A \in \caB(\caH_{0})$, and $f$ is a
function on the spectrum of $A$, then the expression
\beq
f(A)  \quad \textrm{  stands for } \quad  j_0^* f(A)j_0, \eeq
where $j_0$ is the embedding of $\caH_0$ into $\caH$.

We set \beq \bbC_+ := \{ z \in \bbC , \Im z >0 \}, \qquad  \bbC_- := \{ z \in \bbC , \Im z <0 \}. \eeq

\section{Dilations} \label{sec: dilation}

Let $\caE$ be a Hilbert space and let the family $\La_{t \in \bbR^+}$ be a
contractive semigroup on $\caE$:
\begin{equation}  \La_t \La_s =\La_{t+s}, \quad \| \La_t \| \leq 1,  \qquad t,s
\in \bbR^+.    \end{equation}
\begin{definition}

\ben
\item We say that $(\caZ,1_\caE,U_{t\in\bbR})$
is a unitary dilation
 of  $\La_{t \in \bbR^+}$ if
 \begin{romanenumerate}
 \item{$\caZ$ is a Hilbert space and $U_{t\in\bbR} \in \caB(\caZ)$ is a
 1-parameter unitary group, }
 \item{$\caE
\subset \caZ$ and
$1_{\caE}$ is the orthogonal
 projection from $\caZ$ onto $\caE$, }
\item{
 for
all $t \in \bbR^+$
\begin{equation} 1_{\caE} {U}_t 1_{\caE} =\La_t. \end{equation}}
 \end{romanenumerate}
\item{ We call a
dilation $(\caZ,1_\caE,{U}_{ t \in \bbR})$ minimal iff \begin{equation} \big \{
{U}_t \caE\ \big|\ t \in \bbR \big \}^{\mathrm{cl}}=\caZ.
\end{equation}} \een

\end{definition}

\noindent  We have the following theorem due to
\cite{NF}:
\begin{theorem}

\ben
\item{
Every contractive semigroup $\La_{t \in \bbR^+}$ has a minimal
unitary dilation $(\caZ,1_\caE,{U}_{ t \in \bbR})$, unique up
to unitary equivalence. }
\item $1_{\caE}
{U}_t1_{\caE}=\La_{-t}^*$ for $t<0$.
\item
If $\La_{t \in \bbR^+}$ is strongly continuous in $t$, then ${U}_{
t \in \bbR}$ can be chosen to be strongly continuous in $t$.
\een
\end{theorem}
In the following we present a construction of a unitary dilation,
which is well suited for the extended weak coupling limit.

In what follows we assume that
the contractive semigroup $\Lambda_t$ is norm continuous. Hence it
has a generator, denoted $-\i
\Gamma \in \caB(\caE)$, so that $\Lambda_t=\e^{-\i t\Gamma}$.
 Since $\Lambda_t$ is contractive, $-\i \Gamma$ is dissipative:
 \begin{equation}  \Im \Ga  =  \frac{1}{\i 2}(\Gamma - \Gamma^*) \leq
   0.\end{equation}

Let $\frh$ be a Hilbert space.
Set $\caZ_{\res} =L^2(\bbR) \otimes \frh
=L^2(\bbR,\frh)$ and
$\caZ=\caE\oplus \caZ_{\res}$.
 Let $1_\caE$ be the orthogonal projection from
$\caZ$ onto $\caE$.

We define an unbounded linear functional  on $L^2(\bbR)$
 with the domain
$ L^1(\bbR) \cap L^2(\bbR) $, denoted $\langle 1 |$, given by the
obvious prescription
\[
\langle1 | f  = \int_{\bbR} f(x)\d x.
\]
By $| 1 \rangle$,
 we denote the adjoint of $\langle 1 |$ in the
sense of forms. (Note that the adjoint of  $\langle 1 |$ in the
sense of forms is different from the adjoint
 in the sense of
operators, in particular, the latter has a trivial domain).

Introduce the operator
 ${Z}_{\res}$ on $\caZ_{\res}$ as the operator of
multiplication by the variable $x$:
\[
({Z}_{\res}f)(x) =x f(x).
\]
 Let  $\nu \in
\caB(\caE,\frh)$, be an operator
 satisfying the
condition
\begin{equation}\label{condition on A}
  \frac{1}{2\i}( \Gamma - \Gamma^*)= -\pi \nu^* \nu.
\end{equation}


\noindent Put ${W}=  | 1 \rangle  \otimes \nu  $ and
${W}^*=\langle 1 |  \otimes  \nu^*$  and remark that the following
expressions
\begin{equation}\label{qforms}
{W}, \qquad  {W}^* ,\qquad  {W} S {W}^*,   \qquad \textrm{with
} S \in \caB(\caE),
\end{equation}
are well-defined quadratic forms on $ \caD := \caE \oplus \big(
(L^1(\bbR) \cap L^2(\bbR)) \otimes_{\mathrm{ al}} \frh) $,
(where $\otimes_{\mathrm{ al}}$
 denotes the
algebraic tensor product).

%
%
%
%
%
%
%
%
%

Now we combine these objects into something that is a priori a
quadratic form on $\caD$, but turns
out to be a bounded operator. For clarity we will explicitly write
the projections $1_\caE$ onto $\caE$ and $1_{\res}$ onto $\caZ_{\res}$. For
$t \geq 0$, we define
\begin{eqnarray}\nonumber
{U}_t &=& 1_{\res} \e^{ -\i
t{Z}_{\res}}1_{\res}\ +\ 1_\caE\e^{-\i t\Gamma}1_\caE \nonumber\\
&&- \i 1_\caE  \int^t_0 \, \d u \, \e^{-\i (t-u)\Gamma} {W}^*
\e^{-iu{Z}_{\res}}  1_{\res}\nonumber\\&&
 - \i
1_{\res} \int^t_0 \, \d u \, \e^{-\i (t-u){Z}_{\res}}{W} \e^{-\i u\Gamma}1_\caE \nonumber\\
&&-  1_{\res}  \int\limits_{0\leq u_1,u_2,  \, u_1+u_2\leq t} \,
 \d u_1 \d u_2 \, \e^{-
\i u_2{Z}_{\res}} {W} \e^{-\i (t-u_2-u_1)\Gamma}{W}^* \e^{- \i u_1
{Z}_{\res}}1_{\res}, \nonumber \\[5mm]
{U}_{-t} &=& {U}^*_{t}.\label{definition group}
\end{eqnarray}
%
%
%
%
For $z\in\bbC_+$, we define

\begin{eqnarray}
\label{definitionresolvent}
&&Q(z)\
:=\
\left [ \begin{array}{cc} 0 & 0
\\
  0 & (z-{Z}_{\res})^{-1 } \end{array}
  \right]\nonumber\\[2.5ex]
&&+ \left [ \begin{array}{cc} (z-\Gamma)^{-1} &
 (z-\Gamma)^{-1} W^*(z-Z_{\res})^{-1}
\\
 (z-Z_{\res})^{-1}W(z-\Gamma)^{-1} &
 (z-Z_{\res})^{-1}W(z-\Gamma)^{-1}W^*(z-Z_{\res})^{-1} \end{array}
  \right];\nonumber\\[2.5ex]
&&Q(\bar z)\ :=\ Q(z)^*.
\end{eqnarray}

Next we define the following quadratic form on $\caD$,  using the matrix
notation with respect to the decomposition $\caZ =\caE
\oplus \caZ_{\res}$:
\begin{equation} \label{def: Zquad}
 {Z}^+ =
\left [\begin{array}{cc} \Gamma & W^*
\\
 W & {Z}_{\res} \end{array}
  \right],\ \ \
 {Z}^- =
\left [\begin{array}{cc} \Gamma^* & W^*
\\
 W & {Z}_{\res} \end{array}
  \right]
.  \end{equation}
Last,  for $k \in \bbN$, we define 
approximants ${W}_k \in \caB(\caE, L^2(\bbR,\frh))$ to the form ${W}$
\begin{equation}\label{def: Wk}
{W}_k u : =  | 1_{[-k,k]} \rangle \otimes \nu,
\end{equation}\
and approximants $Z_{\res,k} \in \caB(\caZ_{\res})$ for $Z_\res$
\beq
Z_{\res,k} := 1_{[-k,k]}(Z_\res) Z_\res,
\eeq
where $1_{[-k,k]}$ denotes the characteristic function of $[-k,k]$. We set
\begin{equation} \label{def: Zquad1}
 {Z}_k =
\left [ \begin{array}{cc} \Re\Gamma & W_k^*
\\
 W_k & {Z}_{\res,k} \end{array}
  \right]
.  \end{equation}



We have
%
%
%
%
%
%
%
%
%
%
\begin{theorem}\label{group property}
Let  ${U}_t$ be as in (\ref{definition group}), $Q(z)$ as in
(\ref{definitionresolvent}), $Z^\pm$
 as in \eqref{def: Zquad}, and  $Z_k$ as in (\ref{def: Zquad1}),
with $\Gamma$  satisfying condition
(\ref{condition on A}). \ben
 \item{  \label{first item of group property}
The family  $Q(z)$ is the resolvent of a self-adjoint operator
${Z}$, that is, there exists a unique self-adjoint operator $Z$ such that
for all $z \in \bbC\backslash\bbR$
\begin{equation} \label{definition G}
Q(z)=(z-{Z})^{-1}.
\end{equation}
}
\item{ ${U}_t$ extends to a unitary, strongly continuous
one-parameter group in $\caB(\caZ )$ and
\begin{equation} {U}_t =\e^{-\i t {Z}}.\end{equation}
}

\item{  \label{second item of group property}
Fix $\Im z_0>0$.
 $\Dom Z$ consists of vectors $\psi$ of the following form:
\begin{equation}\psi=\left[\begin{array}{c}u\\(z_0-Z_{\res})^{-1}Wu+g\end{array}
\right],\
 \ \
 u\in\caE,\ g\in\Dom Z_{\res},\label{equ}\end{equation}
and $Z$ transforms $\psi$ into
\begin{equation} Z\psi=\left[\begin{array}{c}\Gamma u+W^*g
\\z_0(z_0-Z_{\res})^{-1}Wu+Z_{\res}g\end{array}
\right].\label{equ1}\end{equation}
}
\item For $\psi\in \Dom Z$, we have
\begin{equation}
Z\psi=\lim_{k\to\infty}Z_k\psi.
\label{equ2}\end{equation}

%
%
%
\item{
For $\psi,\psi' \in \caD$, the function $\bbR\ni t
\mapsto \langle \psi| {U}_t  \psi'
\rangle $ is differentiable away from $t=0$, its derivative
$t\mapsto \frac{\d}{\d t} 
 \langle \psi| {U}_t  \psi' \rangle$ is continuous away from $0$ and
at
$t=0$ it has the left and the
right  limit equal respectively to
\begin{equation}
-\i\langle \psi| {Z}^+ \psi' \rangle  =
\lim_{t \downarrow 0}t^{-1}\langle \psi| ({U}_t-1)  \psi' \rangle,
\label{deri1}\end{equation}
\begin{equation}
-\i\langle \psi| {Z}^- \psi' \rangle
 =
\lim_{t \uparrow 0}t^{-1}\langle \psi| ({U}_t-1)  \psi' \rangle.
\label{deri2}\end{equation}
 }

\item{The group ${U}_t$ dilates the semigroup generated by $- \i \Gamma$, that
    is, for
    $ t\geq 0$, \begin{equation}
1_\caE {U}_t 1_\caE= \e^{-\i t \Gamma}.
\end{equation} }
\item{This dilation is minimal iff $\mathfrak{h} =\Ran \nu$.}

%
%
%
%
%
%
%

%
%
%

\een

\end{theorem}
\begin{remark}
Naturally, every densely defined operator gives rise to a
quadratic form on its domain. However, ${Z}^{+}$ and $Z^-$ are
not derived from ${Z}$ in this way.
This is seen from the
explicit description of these domains, as well as from the
fact that for $\psi\in\Dom Z$ we have
$\frac{\d}{\d t}U(t)\psi\Big|_{t=0}=-\i Z\psi$, which should be compared with
(\ref{deri1}) and (\ref{deri2}).
\end{remark}
\begin{remark}
Motivated by
(\ref{def: Zquad1}) and
(\ref{equ2}), we can say that in some sense the operator $Z$ is given by the
matrix
\begin{equation} \label{equ3}
 {Z} =
\left [ \begin{array}{cc} \Re\Gamma & W^*
\\
 W & {Z}_{\res} \end{array}
  \right]
.  \end{equation}
One should however remember, that
strictly speaking the expression  (\ref{equ3})
does not define an operator. To define it an
 appropriate ``renormalization'' is needed: one needs to
 impose a symmetric cutoff and then remove it.
 The precise meaning of this renormalization is described
by (\ref{equ}) and (\ref{equ1}), or by (\ref{equ2}).
Nevertheless,
in the sequel, we will freely use  expressions of the form (\ref{equ3})
remembering that its meaning is given by Theorem
\ref{group property}.
\end{remark}

\begin{remark}\label{rem: scaling invariance}
For $\lambda\in\bbR$, introduce the following unitary operator on
$\caZ$
\[j_\lambda u=u,\ \ u\in\caE;\ \ \ \ \ j_\lambda
g(y):=\lambda^{-1}g(\lambda^{-2}y),\ \ g\in \caZ_{\res}.\]
Note that \[j_\lambda^*Z_{\res}j_\lambda=\lambda^2 Z_{\res},\ \ \
j_\lambda^*|1\rangle=\lambda|1\rangle.\]
Therefore, the operator $Z$ enjoys the following scaling property, which plays
an important role in the extended weak coupling limit:
\[
\lambda^{-2}  j_\lambda^*
\left[\begin{array}{cc} \lambda^2\Re\Gamma & \lambda W^*
\\
 \lambda W & {Z}_{\res} \end{array}
  \right]j_\lambda
=
\left [ \begin{array}{cc} \Re\Gamma & W^*
\\
 W & {Z}_{\res} \end{array}
  \right]
.\]

\end{remark}

\section{Weak coupling limit}\label{stoch}
\subsection{Notation and Assumptions}
 Let $\caE$ and $\caH_{\res}$ be Hilbert spaces. We assume that
$\caE$ is finite dimensional. We set $\caH=\caE\oplus \caH_{\res}$.

 Fix a self-adjoint operator $H_{\res}$ on $\caH_{\res}$ and
 ${E}$ on $\caE $.
 Let the free Hamiltonian $H_0$ on $\caH$ be given as
 \[
 H_0={E} \oplus H_{\res}
. \]
 Let $V \in \caB(\caE,\caH_{\res})$. By a slight abuse of
 notation we denote by $V$ the corresponding operator on $\caH$.
For $\la\in\bbR$, let the interacting Friedrichs Hamiltonian be
 \begin{equation} \label{friedrichs hamiltonian}
 H_{\la}=H_{0}+\la (V+V^*).
 \end{equation}
 We write ${E}=\sum_{e \in \mathrm{sp}{E}} e 1_{e}(E) $ where $e,1_{e}(E)$
 are the eigenvalues and spectral projections of $E$. The spectral subspace of
 ${E}$ for $e$ is denoted $\caE_e$.
 Let us list the assumptions that we will use in our construction.
\\

\noindent{\bf A1}: \emph{Let $\frh_0,\frh_1,\frh_2,\dots,\frh_\infty$
denote the
  Hilbert spaces of dimension $0,1,2,\dots,\infty$. We assume that
there exists a partition of $\bbR$ into  measurable sets
$I_0,I_1,I_2,\dots,I_\infty$ and a unitary identification
\begin{equation}
\caH_{\res}\simeq\int_\bbR^\oplus \frh(x)\d x\simeq
\mathop{\oplus}\limits_{n=0}^\infty
L^2(I_n)\otimes\frh_n,\label{identi}\end{equation}
where $\frh(x):=\frh_n$ for $x\in I_n$, and $H_{\res}$ is the operator of the
multiplication by the variable $x$. Thus, if $f=\int\limits_\bbR^\oplus f(x)\d
x\in\caH_{\res}$, then
\[(H_{\res}f)(x)=xf(x),\]
for Lebesgue almost all $x$.
Moreover, there exists
 a measurable function \[\bbR\ni x\mapsto v(x)\in B(\caE,\frh(x))\]
such that for Lebesgue a.a. $x\in\bbR$ and all $u \in \caE$
\begin{equation}\label{notation V}
(Vu)(x)=v(x)u.
\end{equation}
}

In what follows, the identification
(\ref{identi}) is fixed and will be  used to define the scaling operator
$J_\lambda$.

%
%
%

\noindent
{\bf A2}: \emph{ For any $e\in\sp{E}$, there exists
  $n(e)\in\{0,1,2,\dots\infty\}$ such that $e$ belongs to the interior of
  $I_{n(e)}$.
We will write $\frh_e$ for $\frh_{n(e)}$.
Moreover, we assume that $v$ is continuous at $\sp {E}$,
 so that for $e\in\sp{E}$, we can
unambiguously define $v(e)\in\caB(\caE,\frh_e)$.}
\\

\noindent {\bf A3}: \emph{There is $\de>0$, such that for a certain $c >0$ and for all $ e \in \sp E$,
\beq
   \|v^*(x)v(x)-v^*(e)v(e)\|\leq c|x-e|^\de.      \eeq
We also assume that $ x \mapsto\| v(x)\|$
is bounded.}

\subsection{The reduced weak coupling limit}

In this subsection we describe the reduced weak coupling limit (or the Davies
limit) for Friedrichs Hamiltonians. The Davies limit is usually given in
its time-dependent version described in Theorem
\ref{thm: davies.dyn}. Its stationary form, which comes from \cite{DF1,DF3},
 has some technical advantages over the time dependent version.

In both   theorems about the reduced weak coupling limit
we do not suppose Assumptions {\bf A1, A2} and
{\bf A3}.

\begin{theorem}[Stationary reduced weak coupling limit] Suppose that for $e\in\sp
E$ and $z \in \bbC_+$
\[\lim_{\epsilon\downarrow0}V^*(e+\epsilon z-H_{\res})^{-1}V\]
exists and is independent of $z$. Set
\[\Gamma_e^\st:=\lim_{\epsilon\downarrow 0}
1_{\caE_e} V^*(e+ \epsilon z -H_{\res})^{-1}V1_{\caE_e},\]
\[\Gamma^\st:=\sum_{e\in\sp E}\Gamma_e^\st.\]
Then
\ben\item for $z\in\bbC_+$,
\begin{equation}\label{convergence davies}
 \lim_{\la \rightarrow 0}
1_\caE \big( z-\la^{-2}(H_{\la}-e) \big) ^{-1}1_\caE
= (z-\Gamma_e^\st)^{-1} 1_\caE;
\end{equation}

\item{
for all continuous
functions with compact support $f \in \caC_{\mathrm c}([0,+\infty])$,
\begin{equation}\label{convergence multiF laplace}
\lim_{\la \downarrow 0}  \int_{\bbR^+} \d t f(t)
\e^{\i\la^{-2}t E}1_\caE \e^{- i\la^{-2}tH_{\la}}1_\caE =  \int_{\bbR^+} \d t
f(t) \e^{-\i t\Gamma^\st},
\end{equation}
} \een where all limits are in operator norm.
\label{davies.st}\end{theorem}

\begin{theorem}[Time-dependent reduced weak coupling limit]
Assume that
\[\lim_{t\to\infty}\int_0^t\e^{\i s E}V^*\e^{-\i s H_{\res}}V\d s\]
exists. Set
\[\Gamma_e^\dyn:=
\lim_{t\to\infty}\int_0^t 1_{\caE_e} V^*\e^{-\i s(H_{\res}-e)}V
1_{\caE_e}\d s ,\]
\[\Gamma^\dyn:=\sum_{e\in\sp E}
\Gamma_e^\dyn.\]
Then
\begin{equation}
\lim_{\la \to 0} \sup_{0 \leq t \leq T} \|
 \e^{\i\la^{-2}t {E}} 1_\caE\e^{-\i\la^{-2}tH_{\la}} 1_\caE -    \e^{- \i t \Gamma^\dyn}
\|=0.
\end{equation}
\label{thm: davies.dyn}\end{theorem}

In practice,  $\Gamma^\st$ and $\Gamma^\dyn$ coincide. They will be denoted
simply by $\Gamma$ and called the Davies generator:

\begin{theorem}[Formula for the Davies generator]
Suppose that Assumptions
 {\bf A1, A2} and
{\bf A3} are true.
 Then the assumptions of Theorems \ref{davies.st} and \ref{thm: davies.dyn}
 are true. Moreover, for
 $e \in \mathrm{sp}{E}$ and $z \in \bbC^+$,
\begin{eqnarray}
 - \i \lim_{t \rightarrow +\infty}
\int_0^{t} \d s \,  V^* \e^{- \i s (H_{\res}-e)}V
&=&
 \lim_{\epsilon \downarrow 0} \,
 V^* \big( e+\epsilon z -H_{\res}\big)^{-1} V
\nonumber\\[3mm]
&=&
    \mathrm{P}\int_{\bbR} \d x \, \frac{ v^*(x)v(x)}{x-e} +\i \pi
    v^*(e)v(e).\nonumber
\end{eqnarray}
where $ \mathrm{P}$ denotes the principal value.
Consequently, the stationary and time dependent Davies generator coincide:
\begin{eqnarray}\label{stat and dyn coincide}
\Gamma_e&:=&
\Gamma_e^\dyn=\Gamma_e^\st \\
&=&1_{\caE_e}
\left(
    \mathrm{P}\int_{\bbR} \d x \, \frac{v^*(x)v(x)}{x-e} +\i \pi
    v^*(e)v(e)\right)1_{\caE_e}.\nonumber\end{eqnarray}
\label{limmi}\end{theorem}
\subsection{Asymptotic space and dynamics}\label{sec: asymptot}

Let $e\in\sp E$.
The asymptotic reservoir space and ``total'' space
 corresponding to $e$
 is
\begin{eqnarray*}
\caZ_{\res_e} &:=&  L^2(\bbR)\otimes \frh_e=L^2(\mathbb{R}, \mathfrak{h}_e
),\\
\caZ_e&:=&\caE_e\oplus\caZ_{\res_e}.
     \end{eqnarray*}
 We have the
projections
\[1_{\caE_e}:\caZ_e\to\caE_e,\ \ \
1_{\res_e}:\caZ_e\to\caZ_{\res_e}.\]

Let $Z_{\res_e}$
be the operator of multiplication by the variable in $\bbR$ on $\caZ_{\res_e}$.
We define the map $\nu_e:\caE_e\to\frh_e$
\[ \nu_e :=
v(e)1_{\caE_e},
\]
 Under
the assumptions {\bf A1, A2, A3}, we
define the operator $\Gamma_e$ on $\caE_e$, as in \eqref{stat and dyn coincide}.

Note that $-\pi \nu_e^*\nu_e=\frac{1}{2 \i} (\Gamma_e-\Ga_e)$,
which is the analog of the condition (\ref{condition on A}) for
$Z_{\res_e}$, $\Gamma_e$ and $\nu_e$ for the space
$\caZ_e=\caE_e\oplus\caZ_{{\res}_e}$.
 One can thus apply the procedure of Section \ref{sec: dilation}
 and construct a unitary dilation
 of the semigroup $\e^{-\i t \Gamma_e}$,  as
 defined in (\ref{definition group}).
We will denote this dilation by $\e^{-\i tZ_e}$.

We construct the full
asymptotic  space as a direct sum of independent reservoirs, for
each eigenvalue of $E$:
\begin{eqnarray*}
\frh &:= &\bigoplus_{e \in \mathrm{sp}{E}} \frh_e, \\
\caZ_{\res}& := & \bigoplus_{e \in \mathrm{sp}{E}}
\caZ_{\res_e}= L^2(\mathbb{R}, \frh ),\\
\caZ &:= &
 \bigoplus\limits_{e \in \mathrm{sp}{E}}
\caZ_e=\caE\oplus \caZ_{\res}.\end{eqnarray*}
%
We have the asymptotic reservoir Hamiltonian
\[Z_\res=\bigoplus_{e\in\sp E}Z_{\res_e}.\]
We define the map $\nu:\caE\to\frh$,
\[\nu:=\bigoplus_{e \in \mathrm{sp}{E}} \nu_e
,\]
where we used the decomposition
$\caE=\bigoplus_{e \in \mathrm{sp}{E}} \caE_e$ and $\frh=
\bigoplus_{e \in \mathrm{sp}{E}} \frh_e$.
We also have the operator $\Gamma$ on $\caE$ as defined in Section \ref{sec: dilation}.

Clearly, $Z_{\res}$, $\Gamma$
and $\nu$  satisfy
the condition (\ref{condition on A}).
 One can thus apply the procedure of Section \ref{sec: dilation}
 and construct a unitary dilation $\e^{-\i tZ}$
 of the semigroup $\e^{-\i t \Gamma}$ on $\caZ=\caE\oplus\caZ_{\res}$.

Obviously, everything we constructed commutes with
the orthogonal projections
$
1_{e}  :\caZ   \ra  \caZ_e
$, and we have
\[Z=\bigoplus_{e \in \mathrm{sp}{E}} Z_e.\]

We define the renormalizing hamiltonian $Z_{\mathrm{ren}}$ on $\caZ$:
\begin{equation}
Z_{\mathrm{ren}} := \sum_{e \in \mathrm{sp}{E}} e 1_{e}={E} +
\sum_{e \in \mathrm{sp}{E}} e 1_{{\res}_e}.
\end{equation}

\subsection{Scaling}\label{sec: scaling}

For any $e\in\sp E$, we choose an open set $\tilde I_e$ such that
$e\in\tilde I_e\subset  I_e$ and
 $\tilde I_e$ are mutually
disjoint.
For $\lambda\in\bbR_+$, define the family of contractions
$J_{\la,e}:\caE_e\oplus\caZ_{\res_e}
=\caZ_e \rightarrow \caE \oplus  L^2(\tilde I_e,\frh_e)$, which on $g_e \in
\caZ_{\res_e}$ act as
\begin{equation}
  (J_{\la,e}g_e)(y)=   \left\{ \begin{array}{ll}  \frac{1}{\la}
        g_e(\frac{y-e}{\la^2}),  & \textrm{ if }   y\in \tilde I_e; \\
        0,  & \textrm{ if } y\in\bbR\backslash \tilde I_e. \\ \end{array}
  \right.
\end{equation}
and on $\caE_e$ equals $1_{\caE_e}$.
Note that
\begin{equation}
J^*_{\la,e} J_{\la,e}=1_{\caE_e}\oplus 1_{\lambda^{-2}(\tilde I_e-e)}(Z_{{\res}_e}), \qquad  J_{\la,e}
J_{\la,e}^* =1_{\caE_e}\oplus 1_{\tilde I_e}(H_{\res}).
\end{equation}

For $\psi=\oplus_{e\in\sp E}\psi_e$ we set
\[J_\lambda\psi:=\mathop{\oplus}\limits_{e\in\sp(E)}J_{\la,e}\psi_e.\]
Note that $J_\lambda$ is a partial isometry from $\caZ$ to $\caH$.
\begin{remark}
The precise form of $J_{\la}$ only matters in a neighbourhood of
$\sp {E}$.
For instance, let $\tilde I_e\ni y\mapsto \eta_e(y)$ be increasing
functions differentiable at $e$ and such that $\frac{\d\eta_e}{\d
  y}(e)=1$ for $e\in\sp (E)$. Set
\begin{equation}
  ( J_{\la,e}^\eta g_e)(y)=   \left \{ \begin{array}{ll}  \frac{1}{\la}
        g_e(\frac{\eta_e (y)-\eta_e(e)}{\la^2})  & \textrm{ if }   y\in \tilde I_e, \\
        0  & \textrm{ if } y\in\bbR\backslash\tilde  I_e. \\ \end{array}
  \right.
\end{equation}
Then
all the statements in this paper remain true if one replaces $J_{\la}$
by ${J}_{\la}^\eta$.
\end{remark}

\subsection{Main results}

In this subsection we state the two main results of our paper.
They say that the dynamics generated by
$H_\lambda$ after an appropriate rescaling and renormalization, for a small
coupling approaches the asymptotic dynamics.
Again, we present two versions of the result:  stationary and time-dependent.

\begin{theorem}[Stationary extended weak coupling limit]
\label{thm: stationary}
Assume {\bf A1, A2, A3}. Let $Z_e$ and $Z$  be as defined in Section \ref{sec: asymptot} and let $J_\la$ be as defined in \ref{sec: scaling}.
\ben
\item{
For any $e\in\sp E$ and  $z \in \mathbb{C}^+ $,
\begin{equation}\label{convergence multiF sum}
\lim_{\la \downarrow 0} J_{\la}^* \big( z-\la^{-2}(H_{\la}-e)\big)
^{-1} J_{\la}
 = (z-{Z_e})^{-1}1_e.
\end{equation}
}
\item{
 For all continuous
functions with compact support $f \in \caC_{\mathrm c}([0,+\infty])$,
\begin{equation}
\lim_{\la \downarrow 0}  \int_{\bbR^+} \d t f(t)
\e^{\i\la^{-2}tZ_{\mathrm{ren}}} J_{\la}^* \e^{- \i\la^{-2}tH_{\la}}
J_{\la} =  \int_{\bbR^+} \d t f(t) \e^{-\i t{Z}},
\end{equation}
} \een where all limits are in operator norm.
\end{theorem}

\begin{theorem}[Time-dependent extended weak coupling limit]\label{thm: dynamic}
Assume {\bf A1, A2, A3}. Let $Z_e$ and $Z$  be as defined in
Section \ref{sec: asymptot} and let $J_\la$ be as defined in
Section \ref{sec: scaling}. For all $\psi \in \caZ$ and $t \in
\bbR$,
\begin{equation}\label{eq: conv dynamic}
\lim_{\la \downarrow 0}    \e^{\i
\la^{-2}tZ_{\mathrm{ren}}}J^{*}_{\la} \e^{-\i
\la^{-2}tH_{\la}}J_{\la}\psi = \e^{-\i tZ} \psi.
\end{equation}

\end{theorem}



\begin{remark}
From the proof of Theorem \ref{thm: dynamic}, it follows
immediately that \eqref{eq: conv dynamic} can be stated uniformly
in $t$ on compact intervals, but in weak operator topology. For
all $\psi,\psi' \in \caZ$ and  $0<T <\infty $,
\begin{equation}
\lim_{\la \downarrow 0} \sup_{0 \leq |t| \leq T} \left | \left
\langle \psi' | \, \e^{\i \la^{-2}tZ_{\mathrm{ren}}}J^{*}_{\la}
\e^{-\i \la^{-2}tH_{\la}}J_{\la}\psi - \e^{-\i tZ} \psi  \right
\rangle \right |=0.
\end{equation}

\end{remark}
\begin{remark}
One can also state \eqref{eq: conv dynamic} in the interaction
picture, avoiding the renormalizing hamiltonian $Z_\ren$. For all
$t \in \bbR$ and $\psi \in \caZ$,
 \begin{equation}\label{stoch lim interaction}
\lim_{\la \downarrow 0}   J^{*}_{\la} \e^{\i \la^{-2}t H_0}
\e^{-\i \la^{-2}tH_{\la}}J_{\la} \psi  = \e^{ \i t Z_\res} \e^{-\i
tZ} \psi.
\end{equation}
This is seen most easily by remarking that for all $ t\in \bbR$
and $\psi \in \caZ$, \beq \lim_{\la \downarrow 0} J_\la^* \e^{\i
\la^{-2}t H_0}  J_{\la} \e^{-\i
\la^{-2}tZ_{\mathrm{ren}}}\psi=\e^{\i t Z_\res}\psi. \eeq
\end{remark}

%
%
%
%
%
%
%
%
%
%
\section{Proofs}

\subsection{Proof of Theorem \ref{group property}}
Statement (1) of Theorem \ref{group property} follows by the arguments described
in a slightly different context in
\cite{DF2} (Theorem 2.1). One can take over the proof of \cite{DF2}  almost
verbatim. For completeness, we reproduce an adjusted proof.

Let ${W}_k \in \caB(\caE, L^2(\bbR,\frh))$ for $k \in \bbN$ be
defined as in \eqref{def: Wk}.
Put
\begin{equation}
\Gamma_k(z) := \Re \Gamma+   {W}^*_k  (z-{Z}_{\res})^{-1}  {W}_k.
\end{equation}
 Obviously, the operator
\begin{equation}
{Z}_k :=  \Re \Gamma+  {Z}_{\res} + {W}^*_k  +  {W}_k
\end{equation}
is a well defined self-adjoint operator on
$\textrm{Dom}{Z}_{\res}$ (since it is a bounded perturbation of
${Z}_{\res}$). By the Feshbach formula (see \eqref{fesh}), one checks that the
resolvent $(z-{Z}_k)^{-1}$ is norm convergent to $Q(z)$: It
suffices to remark that for all $z \in \bbC \setminus \bbR$,
\begin{eqnarray}
&&\lim_{k \ra \infty} \Gamma_k(z)=\Gamma(z), \\
&&\lim_{k \ra \infty}  {W}^*_k  (z-{Z}_{\res})^{-1} = {W}^*
(z-{Z}_{\res})^{-1}
\end{eqnarray}
in norm.
It follows that $Q(z)$ satisfies the resolvent formula. To obtain
that $Q(z)$ is actually the resolvent of a (uniquely defined)
self-adjoint operator, it suffices (see \cite{Ka}) to establish for
all $z \in \bbC \setminus \bbR$,
 \ben \item{ \label{kato 1} $\mathrm{Ker}Q(z)=\{0\}$, }
\item{ \label{kato 2} $\mathrm{Ran}Q(z)$ is dense in $\caZ$, }
\item{ \label{kato 3} $Q^*(z)=Q(\bar{z})$. }
\een
\eqref{kato 3} is obvious. To prove \eqref{kato 1}, we let $u \oplus g \in \caE
\oplus \caZ_{\res}$ and we assume $Q(z)u \oplus g=0$. Suppose that e.g.
 $z\in\bbC_+$. Then
\begin{eqnarray}
\label{eq1}
 (z-\Gamma)^{-1}\left(u+W^*(z-Z_{\res})^{-1}g
\right)&=&0,\\\label{eq2}
(z-Z_{\res})^{-1}W(z-\Gamma)^{-1}\left(u+W^*(z-Z_{\res})^{-1}g\right)+(z-Z_{\res})^{-1}g&=&0.
\end{eqnarray}
Inserting (\ref{eq1}) into (\ref{eq2}) yields
$(z-{Z}_{\res})^{-1 }g=0$ and hence $g=0$. Combined with (\ref{eq1}), the latter implies
$u \oplus g=0$.

Using \eqref{kato 1} and \eqref{kato 3}, we get \eqref{kato 2}, since
\begin{equation}
\mathrm{Ran}Q(z)^{\perp} =
\mathrm{Ker}Q(z)^*=\mathrm{Ker} Q(\bar z)=\{0\}.
\end{equation}
Hence, statement 1 of Theorem \ref{group property} is proven.

 To prove Statement (2) we take $\psi,\psi'\in\caD$ and compute the
 following Laplace transform:
\begin{equation}
-\i \int^{+\infty}_{0} \d t  \, \e^{\i zt} \langle\psi|{U}_t\psi'\rangle =
\langle\psi|Q(z)\psi'\rangle.
\label{qq1}
\end{equation}
By functional calculus and the fact that $Q(z)=(z-Z)^{-1}$,
\begin{equation}
-\i \int^{+\infty}_{0} \d t  \, \e^{\i zt} \langle\psi|\e^{-\i t Z}
\psi'\rangle =
\langle\psi|Q(z)\psi'\rangle. \label{qq2}
\end{equation}
Both $t\mapsto \langle\psi|{U}_t\psi'\rangle $ and
$t\mapsto \langle\psi|\e^{-\i tZ}\psi'\rangle$ are continuous functions and we
can apply the inverse Laplace transform to (\ref{qq1}) and (\ref{qq2}), which
yields $\langle\psi|{U}_t\psi'\rangle =\langle\psi|\e^{-\i tZ}\psi'\rangle
$. By the  density of $\caD$ we obtain
${U}_t =\e^{-\i tZ}$. This in particular proves that $U_t$ satisfies the group
property.

To prove Stratement (3) we note that any vector in $\caZ_{\res}$
can be written as $(z_0-Z_{\res})g$ for some $g\in \Dom Z_{\res}$.
Given such $g$, any vector in $\caE$ can be written as
$(z_0-\Gamma)u-W^*g$ (here we use the invertibility of
$z_0-\Gamma$). Set
\[\phi:=\left[\begin{array}{c}(z_0-\Gamma)u-W^*g\\(z_0-Z_{\res})g\end{array}
\right].\] Then $\psi=Q(z_0)\phi$ equals (\ref{equ}) and
$Z\psi=-\phi+z_0Q(z_0)\phi$ equals (\ref{equ1}).

Statements (4)-(6) follow by straightforward calculations.

 To prove Statement (7), we observe that
\begin{equation} \label{span1}
\Span\big \{ e^{-\i t {Z}} \caE\ ,\ t \in \bbR \big
\}^{\mathrm{cl}}=\Span\big \{ (z-{Z})^{-1} \caE\ ,\ z \in \bbC
\setminus \bbR \big \}^{\mathrm{cl}}.
\end{equation}
Since $ \Span\big \{x \mapsto (z-x)^{-1} \ ,\
z \in \bbC \setminus \bbR \big \}$ is dense in $ L^2(\bbR)$, and using
 the fact that $(z-\Ga)^{-1}$,
 $z\in\bbC_+$ and  $(z-\Ga^*)^{-1}$, $z\in\bbC_-$,  are
 invertible, we have
\begin{equation}
\Span\big \{ (z-{Z})^{-1} \caE\ ,\ z \in \bbC \setminus \bbR \big
\}^{\mathrm{cl}}= \{  u \oplus ( L^2(\bbR) \otimes \nu u )  ,\ u \in \caE \big \}.
\end{equation}
This easily implies
 Statement (7).
\qed

%
%
%
%
%
%

\subsection{Proof of Theorem \ref{davies.st}}

Theorem \ref{davies.st} is essentially a special case
of Theorem 3.2 from
\cite{DF3} (see also \cite{DF1}. For the convenience of the reader, and because the case we
consider allows for some
 simplifications, we sketch the proof below.

Let \begin{equation} G^{-1}(e,\la,z) := 1_\caE \big(
z-\la^{-2}(H_{\la}-e)\big)^{-1} 1_\caE,
\end{equation}
which yields immediately the bound
\begin{equation}\label{bound}
\| G^{-1}(e,\la,z) \| \leq |\Im z|^{-1}.
\end{equation}
In the following we simplify the notation $G(e,\la,z)$ into
$G$ (hence, we fix a certain $e \in \sp E$) and we put
\begin{equation}
G_\d= \sum_{e' \in \sp E} 1_{\caE_{e'}} G 1_{\caE_{e'}},
\qquad G_\off := G-G_\d
\end{equation}
and $1_{\caE_{\underline{e}}} := 1_\caE-
1_{\caE_{e}}$.\\
By the Feshbach formula (see futher: \eqref{fesh}), we have
\begin{equation}\label{feshbach1}
G= z-\la^{-2}({E}-e)
 -\la^{-2} 1_\caE V^*\left(z-\la^{-2}(H_{\res}-e)\right)^{-1}V1_\caE. \nonumber
\end{equation}
By the assumption of Theorem \ref{davies.st}, it is immediate that
\begin{equation}\label{conv diagonal 1}
\lim_{\la \downarrow 0}1_{\caE_{e}}G_\d^{-1}=
(z-\Gamma_e^\st)^{-1}.
\end{equation}
By the Neumann expansion and the assumption of Theorem
\ref{davies.st}, one has for small enough $\la$ and some $c>0$,
\begin{equation} \label{conv diagonal 2}
\| 1_{\caE_{\underline{e}}} G^{-1}_d \| \leq     c \la^{2}, \qquad
\|G_\off \| < c.
\end{equation}
\noindent From $G=G_\d+G_\off$, we deduce
\begin{equation}
 G^{-1}=G_\d^{-1}-G_\d^{-1}G_\off G_\d^{-1}+G_\d^{-1}G_\off G_\d^{-1}G_\off G^{-1},
\end{equation}
from which
\begin{equation} \label{vanishing nondiag 1}
 1_{\caE_e}(G^{-1}-G_\d^{-1})=- 1_{\caE_e}G_\d^{-1}  G_\off 1_{\caE_{\underline{e}}}G_\d^{-1} (1    -   G_\off G^{-1} ).
\end{equation}
Using \eqref{bound}, \eqref{conv diagonal 1} and (\ref{conv diagonal 2}), we
see that the
right hand side of \eqref{vanishing nondiag 1} vanishes, yielding
\begin{equation}\label{conv diagonal 1a}
\lim_{\la \downarrow 0}1_{\caE_{e}}G^{-1}= \lim_{\la
\downarrow 0}1_{\caE_{e}}G_\d^{-1}
=(z-\Gamma_e^\st)^{-1}.
\end{equation}
Writing
\begin{equation} \label{vanishing nondiag 3}
 1_{\caE_{\underline{e}}}G^{-1}= 1_{\caE_{\underline{e}}} G_\d^{-1}-  1_{\caE_{\underline{e}}} G_\d^{-1} G_\off G^{-1}
\end{equation}
and using \eqref{conv diagonal 2}, one sees that
\begin{equation} \label{vanishing nondiag 4}
\lim_{\la  \downarrow 0} 1_{\caE_{\underline{e}}}G^{-1}=0.
\end{equation}
Together, \eqref{conv diagonal 1a} and \eqref{vanishing nondiag
4} end the proof of (1).

(2) follows from (1) as in \cite{DF1}.
\qed

\subsection{Proof of Theorem \ref{thm: davies.dyn}}

 Theorem \ref{thm: davies.dyn} is a special case of a well known result of Davies
 \cite{Da1}, reproduced e.g. in \cite{DF1}.
For the convenience of the reader, and because some
simplifications are possible, we
 sketch the proof below.

We start from the following representation for
$\La_{t,\la} :=  \e^{ \i t \la^{-2}  {E}} 1_\caE  \e^{- \i t
\la^{-2} H_{\la}} 1_{\caE}$:
\begin{equation} \La_{t,\la}=1 +  \int^{t}_0  D_{\la,t}(u)
\La_{u,\la} \d u,
 \end{equation}
with
\begin{eqnarray}
D_{\la,t}(u) &=& \  \la^{-2} \int^{t}_{u}   e^{\i
\la^{-2}v {E}}  V^*\e^{-\i \la^{-2}
(v-u) H_{\res}}  V \e^{-\i \la^{-2} u {E}}  \d v  \\
&=& \sum_{e,e' \in \sp E}  \int^{\la^{-2}(t-u)}_0   1_{\caE_e} V^*\e^{-\i
s (H_{\res}-e)}  V 1_{\caE_{e'}} \e^{-\i \la^{-2} u (e'-e)}  \d s. \nonumber
 \end{eqnarray}

\noindent Let for $T >0$, ${Q} := \caC_0([0,T])$ be the Banach
space of continuous functions, equipped with the supremum norm.
Define the operators $K_{\la}$ and $K$ on  ${Q}$ by (for $0 \leq t
\leq T$)
\begin{equation}
(K_{\la} f)(t)= \int_0^t   D_{\la,t}(s) f(s) \d s, \qquad   (K
f)(t)= -\i \Gamma^{\mathrm{dyn}} \int_0^t     f(s)  \d s.
 \end{equation}
We will prove that
\begin{equation}  \label{convergence volterra} \mathrm{s}-\lim_{\la
\downarrow 0} K_{\la}=K.
\end{equation}
Let
\begin{equation}
 \tilde{\Gamma} :=- \i \lim_{t \rightarrow
+\infty} \int_0^{t}   V^* \e^{- \i s
(H_{\res}-e)}V  \d s,
\end{equation}
whose existence was proven in Theorem \ref{limmi}.

One checks that for all $ t \in [0,T]$
\begin{equation} \label{proof volterra 1}
\lim_{\la \downarrow 0} \left | (K_{\la} f)(t) + \i  \sum_{e,e'}
\int_0^t   1_{\caE_e}\tilde{\Gamma} 1_{\caE_{e'}} \e^{-\i \la^{-2}
s (e'-e)} f(s) \d s \right |=0,
 \end{equation}
which follows by the assumption of Theorem \ref{thm: davies.dyn} and dominated convergence. Since
$f$ is (bounded and continuous, hence) integrable,
the Riemann-Lesbegue lemma yields, for $e,e' \in \sp E$,
 \begin{equation} \lim_{\la \downarrow 0}\i
\int_0^{t}  1_{\caE_e}\tilde{\Gamma} 1_{\caE_{e'}} \e^{-\i \la^{-2} s
(e-e')} f(s) \d s =  \de_{e,e'} \int_0^{t}
 \Gamma_e^{\mathrm{dyn}}   f(s) \d s, \end{equation}
and hence \eqref{proof volterra 1} proves
(\ref{convergence volterra}).
 Note that $\La_{t,\la}$ and $\La_t :=
\e^{-\i t \Gamma^{\mathrm{dyn}}}$ satisfy the equations.
\begin{equation} \label{integral equations}
\La_{\la}=\La_0 +K_{\la} \La_{\la},  \qquad \La=\La_0+K \La,
 \end{equation}
where $\La_0$ is the constant function with value $\La_0
=\La_{0,\la}=1$. Remark that by the assumption of Theorem \ref{thm: davies.dyn}, there exists a constant
$c$ and a $\lambda_0$ such that for all $\la \leq \la_0$ and for
all $n \in \bbN$, we have
\begin{equation} \label{bound volterra}
 \|K_{\la}^n \| \leq \frac{(ct)^n }{n!},
 \qquad   \|K^n \| \leq \frac{(ct)^n }{n!}.
 \end{equation}
This means that both
\begin{equation}
(1-K_{\la})^{-1}= \sum_{n=0}^{+\infty} K^n_{\la},  \qquad
(1-K)^{-1}= \sum_{n=0}^{+\infty} K^n
 \end{equation}
exist and that for each $n \in \bbN$, $ \mathrm{s}-\lim_{\la
\downarrow 0} K^n_{\la}=K$. By \eqref{integral equations}, we thus
have
\begin{equation}
\La_{\la}-\La =((1-K_{\la})^{-1} - (1-K)^{-1} )\La_0=
\sum_{n=0}^{+\infty} \big( K^n_{\la}- K^n_{\la} \big) \La_0
. \end{equation}
Since each term in the right-hand side vanishes as  $\la
\downarrow 0$ and the sequence is absolutely convergent by
\eqref{bound volterra}, Theorem \ref{thm: davies.dyn} follows.\\

\subsection{Proof of Theorem \ref{limmi}}

Let us first state a general lemma about the principal value.

\begin{lemma}
 Let $f$ be a bounded function on $\bbR$ such that $\frac{f}{1+|x|}\in L^1(\bbR)$,
$f$ is continuous at $0$ and there exist $\delta,C>0$ such that
for $|x|<C \Rightarrow |f(x)-f(0)|\leq |x|^\delta$.
Then, for $z \in \bbC_+$,
\begin{eqnarray}
 - \i \lim_{T \rightarrow +\infty}
\int_0^{+T} \d t \,  \int_{\bbR} \d x f(x) \e^{- \i t x} &=&
 \lim_{\epsilon \downarrow 0} \int_\bbR
f(x) \big( \epsilon z -x\big)^{-1} \d x
\label{expre}\\[3mm]
&=&
-    \mathrm{P}\int_{\bbR}  \frac{f(x)}{x}\d x  +\i \pi
    f(0).\nonumber
\end{eqnarray}
\end{lemma}

\proof For the first expression of (\ref{expre}), we write
\begin{eqnarray}
&& - \i \lim_{T \rightarrow +\infty}
\int_0^{+T} \d t   \int_\bbR \,\d x f(x) \e^{- \i t x} \nonumber\\
&=&\int_\bbR \frac{(-1+\e^{-\i Tx})}{x} f(x) \d x \nonumber\\[3mm]
&=&f(0)\int_{|x|\leq C}\frac{-1+\e^{-\i Tx}}{x}\d x \nonumber\\[3mm]&&
+\int_{|x|\leq C}\frac{-f(x)+f(0)}{x}\d x
+\int_{|x|\leq C}\frac{(f(x)-f(0))\e^{-\i Tx}}{x}\d x \nonumber
\\[3mm]
&&-\int_{|x|>C}\frac{f(x)}{x}\d x +\int_{|x|>C}\frac{f(x)\e^{-\i Tx}}{x}\d x
.\label{coco}\end{eqnarray}
 The first term, by the residue calculus
goes to $f(0)\i\pi$.
By the Riemann-Lebesgue Lemma, the third and the fifth term on the right of
(\ref{coco}) go to zero. The second and fourth term yield
$\mathrm{P}\int\frac{f(x)}{x}\d x$.

To get the second equality in \eqref{expre}, we write
$z=a+\i b$ and compute:
\begin{eqnarray}
&&  \int_\bbR
f(x) \big( \epsilon z -x\big)^{-1} \d x  \nonumber\\[3mm]
&=&  \int_\bbR
 \frac{\epsilon \i b  f(x) }{  (\epsilon a -x)^{2} +(\epsilon b)^2 }\d x
 \nonumber
\\
&&+ \int_{|x|<\mu} f(x)\left( \frac{ (\epsilon a-x) }{  (\epsilon a -x)^{2} +
(\epsilon b)^2 }-\frac{-x}{x^2+(\epsilon b)^2}\right)\d x\nonumber\\&&
- \int_{|x|<\mu}  \frac{x f(x)}{  x^{2} +
(\epsilon b)^2 }\d x
+ \int_{|x|>\mu}  \frac{ (\epsilon a-x) f(x)}{  (\epsilon a -x)^{2} +
(\epsilon b)^2 }\d x,\label{splitting in lemma}
\end{eqnarray}
where $0<\mu<1$ is  fixed.
The sum of the last two terms converges to
$-\mathrm{P}\int\limits_{\bbR}  \frac{ f(x)}{  x }\d x$.
The second term can be estimated by
\begin{eqnarray*}&&\sup|f|\int_{|x|<\mu}\left(
\frac{|\epsilon a-x|}{(x-\epsilon a)^2+\epsilon^2 b^2}-
\frac{|x|}{x^2+\epsilon^2 b^2}\right)\d x
\\&=&
\frac{\sup|f|}{2}\left|\log
\frac{((\mu+\epsilon a)^2+\epsilon^2 b^2)((\mu-\epsilon a)^2+\epsilon^2 b^2)}
{(\mu^2+\epsilon^2 b^2)^2}\right|\underset{\epsilon\to0}{\to}0.
\end{eqnarray*}
\qed

To apply this lemma, it suffices to note that
$f(x):=v^*(x)v(x)$ is a bounded $L^1$ function, continuous and H\"older at $\sp E$.

%
%
%
%
%
%
\subsection{Proof of Theorem \ref{thm: stationary}}

\begin{lemma}\label{lem: stat}
Let $e,e'\in\sp E$  and $z\in\bbC_+$. Then
\[\lim_{\lambda\downarrow0}
\frac{1}{\lambda}V^*
(z-\lambda^{-2}(H_{\res}-e))^{-1}J_{\lambda,e'}
=\langle 1|{\otimes} \ v^*(e) \ (z-Z_{{\res}_e})^{-1}\delta_{e,e'}.\]
 \end{lemma}

\proof
Let $g_{e'}\in\caZ_{\res_{e'}}$. Then
\begin{eqnarray}&&
\frac{1}{\lambda}V^*
(z-\lambda^{-2}(H_{\res}-e))^{-1}J_{\lambda,e'}g\nonumber\\
& =&
\frac{1}{\lambda^2}\int_{\tilde I_{e'}}
v^*(y)\left(z-\frac{y-e}{\lambda^2}\right)^{-1}g_{e'}
(\frac{y-e'}{\lambda^2})\d y
\nonumber\\
&=&
\int\limits_{\lambda^{-2}(\tilde  I_{e'}-e')}
v^*(e'+\lambda^2 x)(z-x+\lambda^{-2}(e-e'))^{-1}g_{e'}(x)\d x.\nonumber
\nonumber\end{eqnarray}
For $e\neq e'$, we estimate the square of the norm by
\begin{eqnarray*}
&&\int\limits_{\lambda^{-2}(\tilde  I_{e'}-e')}
\|v^*(e'+\lambda^2 x)(z-x+\lambda^{-2}(e-e'))^{-1}\|^2\d x
\int_{\bbR}\|g_{e'}(x)\|^2\d x\\
&\leq&\mathop{\sup}\limits_{y \in \bbR}\|v(y)\|^2\int\limits_{\lambda^{-2}(\tilde  I_{e'}-e')}
|(z-x+\lambda^{-2}(e-e'))^{-1}|^2\d x\int_{\bbR}\|g_{e'}(x)\|^2\d x\to 0.
\end{eqnarray*}
The first integral in the last line vanishes by Lesbegue dominated convergence since $e \notin \left( I_{\e'}-e' \right)$.
For $e=e'$,
\begin{eqnarray*}
&&\left\|\int\limits_{\lambda^{-2}(\tilde  I_{e}-e)}
\left(v^*(e+\lambda^2 x)-v^*(e)\right)(z-x)^{-1}g_{e}(x)\d x\right\|^2
\\&\leq&\int_{\bbR}
\left\|\left(v^*(e+\lambda^2x)-v^*(e)\right)(z-x)^{-1}\right\|^2\d x\int_{\bbR}
\|g_e(x)\|^2\d x\to0
,\end{eqnarray*}
by the Lebesgue dominated convergence theorem, since $v$ is bounded and continuous in $e$.
Since $g_{e'}$ enters the above
estimates
 only via $  \| g_{e'}\|^2 =
 \int_{\bbR}     \| g_{e'}(x)\|^2 \d x $, the convergence is in norm.
\qed
\medskip

The proof of Theorem \ref{thm: stationary}  is based on the formula
\begin{eqnarray}
(z-H_\lambda)^{-1}&=&(z-H_{\res})^{-1}\nonumber\\&&+
\left(1_\caE+\lambda(z-H_{\res})^{-1}V \right)
G_\lambda(z)\nonumber\\&&\times \left(1_\caE+\lambda
V^*(z-H_{\res})^{-1}\right), \label{fesh}\end{eqnarray} where
$G_\lambda(z):=1_\caE(z-H_\lambda)^{-1}1_\caE$. After appropriate
rescaling and sandwiching with $J_{\lambda,e'}^*$ and
$J_{\lambda,e''}$, (\ref{fesh}) becomes
\begin{eqnarray}
&&J_{\lambda,e'}^*(z-\lambda^{-2}(H_\lambda-e))^{-1}J_{\lambda,e''}\nonumber\\
&=&\delta_{e',e''}1_{\lambda^{-2}(\tilde
  I_{e'}-e')}(Z_{\res,e'})(z-Z_{\res,e'}-\lambda^{-2}(e'-e))^{-1}
\nonumber\\
&&+\left(1_{\caE_{e'}}+J_{\lambda,e'}^*\frac1\lambda(z-
\lambda^{-2}(H_\lambda-e))^{-1}V\right)\nonumber\\
&&\times\ \  G_\lambda(z,e)\nonumber
\\
&&
\times\left(V^*(z-\lambda^{-2}(H_\lambda-e))^{-1}\frac1\lambda J_{\lambda,e''}
+1_{\caE_{e''}}\right),
\label{nonu}\end{eqnarray}
where
\[G_\lambda(z,e):=
 1_{\caE}
(z-\lambda^{-2}(H_\lambda-e))^{-1}1_{\caE}.\]
The first term of (\ref{nonu})  has $\delta_{e',e''}$, because $\tilde
I_{e'}$ and   $\tilde I_{e''}$ are disjoint. It converges to
\[\delta_{e,e'}\delta_{e,e''}(z-Z_{{\res}_e})^{-1}.\]
By the stationary Davies limit (Theorem \ref{davies.st} (1)),
\[G_\lambda(z,e)\to 1_{\caE_e}(z-\Gamma_e)^{-1}1_{\caE_e}.\]
Finally, by application of Lemma \ref{lem: stat},
the second term of the rhs of (\ref{nonu}) converges to
\begin{eqnarray}
&&\delta_{e,e'}\delta_{e,e''}\left(1_{\caE_{e}}+
(z-Z_{\res,e})^{-1} \,
| 1\rangle \otimes v(e) \right)\nonumber\\
&&\times\ \  1_{\caE_e}(z-\Gamma_e)^{-1} 1_{\caE_e}\nonumber
\\
&&
\times\left(\langle 1| \otimes v^*(e) \, (z-Z_{\res,e})^{-1}
+1_{\caE_{e}}\right).
\nonumber\end{eqnarray}
\qed

\subsection{Proof of Theorem \ref{thm: dynamic} }

We start with the time dependent analog of Lemma
\ref{lem: stat}.

\begin{lemma} Let $g_e\in L^1(\bbR,\frh_e) \cap L^2(\bbR,\frh_e)  =
  \caD\cap\caZ_{{\res},e}$. Then, uniformly for $|t|<T$, we have the convergence
\[\lambda^{-1}V^*\e^{\i t\lambda^{-2}(e-H_{\res})}
J_{\lambda,e}g_e\to
\langle 1|\otimes v^*(e)\e^{-\i tZ}g_e.\]
\end{lemma}

\proof
\begin{eqnarray}
\frac{1}{\lambda}V^*
\e^{-\i t\lambda^{-2}(H_{\res}-e)}J_{\lambda,e'}g_e
& =&
\frac{1}{\lambda^2}\int_{\tilde I_{e}}
v^*(y)\e^{-\i t\lambda^{-2}(y-e)}g_{e}
(\frac{y-e'}{\lambda^2})\d y
\nonumber\\
&\!\!\!\!\!\!\!\!\!\!=&\!\!\!\!\!\!\!\!\!
\int\limits_{\lambda^{-2}(\tilde  I_{e}-e)}
v^*(e+\lambda^2 x)\e^{\i t x}g_{e}(x)\d x\nonumber\\&
\to& v^*(e)\int \e^{-\i tx}g(x)\d x.
\nonumber\end{eqnarray}
\qed

The proof of Theorem
\ref{thm: dynamic}   is based on the time dependent analog of
the
formula (\ref{fesh}):
\begin{eqnarray} \label{eq: dyson}
\e^{-\i t H_\lambda}&=
&\e^{-\i t H_{\res}}
\ +\ T_{\la}(t) \nonumber\\ &&+
\i\lambda\int_0^t T_\lambda(t-s)
V^*\e^{-\i sH_{\res}}\d s +
\i\lambda\int_0^t
\e^{-\i sH_{\res}}V
T_\lambda(t-s)
\d s\nonumber\\
&&-\lambda^2\mathop{\int\int}\limits_{0\leq s_1,s_2,\
s_1+s_2\leq  t}
\e^{-\i s_1 H_{\res}}V
T_\lambda(t-s_1-s_2) V^*
\e^{-\i s_2H_{\res}}\d s_1\d s_2 ,\nonumber
\end{eqnarray}
where
\[T_\lambda(t):=1_\caE\e^{-\i tH_\lambda}1_\caE.\]

Rescaling, multiplying from the left by $\e^{\i t\lambda^{-2} Z_\ren}
J^*_{\la,e}$ and from the right by $J_{\la,e'}$,
 we obtain
\begin{eqnarray} \label{eq: dyson sandwich}
&&1_{e}\e^{\i\la^{-2}t Z_\ren} J^*_{\la}\e^{-\i \la^{-2} t
    H_\lambda}J_{\la}1_{e'}
\nonumber\\
&=& J^*_{\la,e}\e^{-\i \la^{-2}t (H_{\res}-e)}  J_{\la,e'}\nonumber\\
&&+ \e^{\i t\lambda^{-2}e} 1_{\caE_e} T_\lambda(t) 1_{\caE_{e'}}\nonumber\\ &&+
\frac{\i}{\lambda} \e^{\i t\lambda^{-2}e}
1_{\caE_e}   \int_0^t   T_\lambda(t-s)
V^*\e^{-\i \la^{-2}s H_{\res}} J_{\la,e'} \d s   \nonumber\\
&&+
\frac{\i}{\lambda} \e^{\i t\lambda^{-2}e}  \int_0^t
J^*_{\la,e} \e^{-\i \la^{-2} s H_{\res}}V
 T_\lambda(t-s)
\d s   \,  1_{\caE_{e'}}  \nonumber\\
&&-\lambda^{-2}
\e^{\i t\lambda^{-2}e}  \mathop{\int\int}\limits_{0 \leq
s_1, s_2,\ s_1+s_2\leq  t}
J^*_{\la,e} \e^{-\i s_1\la^{-2} H_{\res}} \nonumber \\ && \times V
T_\lambda(t-s_1-s_2) V^*
\e^{-\i \la^{-2} s_2 H_{\res}} J_{\la,e'}  \d s_1\d s_2,
\label{nonu1}
\end{eqnarray}
where
\[T_\lambda(t):=1_\caE   \e^{-\i \la^{-2}t
  H_\lambda}1_\caE.\]

The first term  of (\ref{nonu1}) converges to
\begin{eqnarray}
\delta_{e,e'}\e^{-\i t Z_{\res}}1_{{\res}_e}.
\end{eqnarray}

To handle the next terms we use repeatedly the fact that
\[\|T_\lambda(s)-\e^{\i
  s\lambda^{-2}E}\e^{-\i s\Gamma}\|\underset{\lambda}\to0\]
uniformly for $0\leq s\leq t$.
The
second term
converges to
\begin{eqnarray*}
\e^{\i t\lambda^{-2}e} 1_{\caE_e} \e^{-\i t\lambda^{-2}E}\e^{-\i t\Gamma}
 1_{\caE_{e'}}
&=&\delta_{e,e'}1_{\caE_e}\e^{-\i t\Gamma}.\end{eqnarray*}

The third term acting on $g_{e'}\in  L^1(\bbR,\frh_e) \cap L^2(\bbR,\frh_e) $, converges to
\begin{eqnarray}&&
\i \e^{\i t\lambda^{-2}e}
1_{\caE_e}   \int_0^t  \e^{-\i( t-s)\lambda^{-2}E}\e^{-\i(t-s)\Gamma}
\langle 1|v^*(e')
\e^{-\i s (Z_{\res}+\lambda^{-2}e')}
 g_{e'}\d s\nonumber  \\
&=&
\i 1_{\caE_e}   \int_0^t \e^{-\i(t-s)\Gamma}
\langle 1|v^*(e')
\e^{-\i s Z_{\res}} g_{e'}\e^{\i s\lambda^{-2}(e-e')}\d s
.\label{nonu2}\end{eqnarray}
If $e-e'\neq0$, this goes to zero by the Lebesgue-Riemann Lemma.
Therefore, (\ref{nonu2}) equals
\begin{eqnarray}&&\delta_{e,e'}
1_{\caE_e}   \int_0^t \e^{-\i(t-s)\Gamma}
\langle 1|v^*(e')
\e^{-\i s Z_{\res}} g_{e}\d s  .\end{eqnarray}

The fourth
 term sandwiched between $g_e\in  L^1(\bbR,\frh_e) \cap L^2(\bbR,\frh_e) $ and $u\in\caE$ converges to
\begin{eqnarray}&&
\i  \int_0^t\left\langle g_e\Big|
\e^{-\i  s Z_{\res}}|1\rangle{\otimes}
v(e)\e^{-\i(t-s)\Gamma}\e^{-\i\lambda^{-2}(t-s)E}
1_{\caE_{e'}}u\right\rangle\e^{\i(t-s)\lambda^{-2}e} \d s\nonumber\\
&=&
\i  \int_0^t\left\langle g_e\Big|
\e^{-\i  s Z_{\res}}|1\rangle{\otimes}
v(e)\e^{-\i(t-s)\Gamma}
1_{\caE_{e'}}u\right\rangle\e^{\i(t-s)\lambda^{-2}(e-e')}\d s.\label{fourth}
\end{eqnarray}
Again, if $e-e'\neq0$, this goes to zero by the Lebesgue-Riemann Lemma.
Therefore, (\ref{fourth}) equals
\begin{eqnarray}\delta_{e,e'}
\i  \int_0^t\left\langle g_e\Big|
\e^{-\i  s Z_{\res}}|1\rangle{\otimes}
v(e)\e^{-\i(t-s)\Gamma}
1_{\caE_{e}}u\right\rangle\d s.
\end{eqnarray}

The fifth term sandwiched between $g_e\in  L^1(\bbR,\frh_e) \cap L^2(\bbR,\frh_e)$
 and $g_{e'}\in  L^1(\bbR,\frh_{e'}) \cap L^2(\bbR,\frh_{e'}) $ converges to
\begin{eqnarray}&&
-  \mathop{\int\int}
\limits_{0 \leq
s_1, s_2,\ s_1+s_2\leq  t}\e^{\i (t-s_1)\lambda^{-2}e}
\left\langle g_e\Big|
 \e^{-\i s_1 Z_{\res}} v(e){\otimes}|1\rangle
 \right.\nonumber \\ &&\times \left.
\e^{-\i(t-s_1-s_2)\lambda^{-2}E}\e^{-\i(t-s_1-s_2)\Gamma} v(e')^*{\otimes}\langle1|
\e^{-\i s_2 (Z_{\res}+\lambda^{-2}e')} g_{e'}\right\rangle  \d s_1\d s_2 \nonumber \\
&=&
-  \sum\limits_{e''\in\sp E}\mathop{\int\int}
\limits_{0 \leq
s_1, s_2,\ s_1+s_2\leq  t}\e^{\i (t-s_1)\lambda^{-2}(e-e'')-s_2\lambda^{-2}
(e'-e'')}
\left\langle g_e\Big|
 \e^{-\i s_1 Z_{\res}}v(e){\otimes}|1\rangle
 \right.\nonumber \\ && \times
\e^{-\i(t-s_1-s_2)\Gamma}1_{\caE_{e''}} v(e')^*{\otimes}\langle1|
\e^{-\i s_2 Z_{\res}} g_{e'}\Big\rangle  \d s_1\d s_2 \label{fifth}.
\end{eqnarray}
By the Riemann-Lebesgue Lemma the terms with $e-e''\neq0$ or $e'-e''\neq0$ go
to zero. Thus (\ref{fifth}) equals
\begin{eqnarray}&&
-\delta_{e,e'}
 \mathop{\int\int}
\limits_{0 \leq
s_1, s_2,\ s_1+s_2\leq  t}
\left\langle g_e\Big|
 \e^{-\i s_1 Z_{\res}} v(e){\otimes}|1\rangle
 \right.\nonumber \\ &&\times
\e^{-\i(t-s_1-s_2)\Gamma}1_{\caE_{e}} v(e)^*{\otimes}\langle1|
\e^{-\i s_2 Z_{\res}} g_{e}\Big\rangle  \d s_1\d s_2 \nonumber.
\end{eqnarray}

Thus we proved that for $\psi,\psi'\in\caD$ we have
\[\sup_{0\leq t\leq T}
\left|\left\langle \psi|
\e^{\i t\lambda^{-2}Z_\ren}
J_\lambda^*\e^{-\i t\lambda^{-2} H_\lambda} J_\lambda
  \psi'\right\rangle-
\left\langle \psi|
\e^{-\i t Z}
  \psi'\right\rangle\right|\underset{\lambda\to0}\to0.\]
By density, this can be extended to the whole
$\caZ$. Using the fact that $\e^{-\i tZ}$ is unitary and
$\e^{\i t\lambda^{-2}Z_\ren}
J_\lambda^*\e^{-\i t\lambda^{-2} H_\lambda} J_\lambda$
contractive we obtain that for $\psi\in\caZ$
\[ \lim_{\la \downarrow 0}
\e^{\i t\lambda^{-2}Z_\ren}
J_\lambda^*\e^{-\i t\lambda^{-2} H_\lambda} J_\lambda
  \psi =
\e^{-\i t Z}
  \psi. \]


%
%
%
%
%
%
%

\medskip

\noindent{\bf Acknowledgments.}
The research of J.~D. was  partly supported by the
Postdoctoral Training Program HPRN-CT-2002-0277 and the Polish grants
 SPUB127 and 2 P03A 027 25. Part of the work was done  when both authors
 visited the
 Erwin Schr\"odinger Institute (J.~D. as a Senior Research Fellow), as well as
 during a visit of
J.~D.  at  K.~U.~Leuven supported by a grant of the ESF. W.~D.~R.
is an Aspirant of the FWO-Vlaanderen.

\bibliographystyle{plain}

\begin{thebibliography}{10}

\bibitem[AFL] {AFL} L.~Accardi, A.~Frigerio, Y.G.~Lu:
 Weak coupling limit as a quantum functional central limit
 theorem,
  \emph{Comm. Math. Phys.} {\bf 131}, 537--570 (1990).

\bibitem[AJPP]{AJPP} W.~Aschbacher, V.~Jak${\check {\mathrm s}}$i\'c, Y.~Pautrat, C.~-A.~Pillet: Introduction to non-equilibrium quantum
statistical
 mechanics, "Open Quantum Systems III Recent Developments" Lecture Notes in
 Mathematics 1882
 eds S. Attal, A. Joye, C.-A. Pillet 
2006

\bibitem[Da1]{Da1} E.~B.~Davies: Markovian master equations,
\emph{Comm. Math. Phys.} {\bf 39}, 91 (1974).

\bibitem[DD]{DD} J.~Derezi\'{n}ski, W.~De~Roeck: Extended weak coupling limit
  for Pauli-Fierz operators, in preparation

\bibitem[DF1]{DF1}  J.~Derezi\'{n}ski, R.~Fr\"uboes: Fermi Golden Rule and
  open quantum systems,
 "Open Quantum Systems III Recent Developments" Lecture Notes in
 Mathematics 1882
 eds S. Attal, A. Joye, C.-A. Pillet 
2006
\bibitem[DF2]{DF2}  J.~Derezi\'{n}ski, R.~Fr\"uboes: Renormalization of Friedrichs Hamiltonians, \emph{Reports on Math. Phys.} 50, 433--438 (2002)

\bibitem[DF3]{DF3}  J.~Derezi\'{n}ski, R.~Fr\"uboes: Stationary van Hove
  limit,  Journ. Math. Phys 46 (2005) 063511

\bibitem[HP]{HP} R.~L.~Hudson, K.~R.~Parthasaraty: Quantum Ito's formula and stochastic evolutions,  \emph{Comm.
Math. Phys.} {\bf 93}  no.~ 3, 301--323 (1984).

\bibitem[Ka]{Ka} T.~Kato:  Perturbation Theory for Linear Operators, 2d ed. Springer-Verlag, Berlin
(1984).

\bibitem[Ku]{Ku} B.~K\"ummerer, W.~Schr\"oder:  A new construction of unitary dilations: singular coupling to white noise, in Quantum Probabilty and Applications, eds L.~Accardi and W.~von Waldenfels, (1984)


\bibitem[NF]{NF} B.~Sz.~Nagy and C.~Foias: Harmonic Analysis of Operators in Hilbert Space, North-Holland, New York (1970)


\bibitem[VH]{VH} L.~Van Hove: Quantum-mechanical
perturbations giving rise to a statistical transport equation.
\emph{Physica} {\bf 21}, 517 (1955).

\end{thebibliography}

 \end{document}